\documentclass[3p]{elsarticle}
\usepackage{lineno,hyperref}
\usepackage{amsmath}
\usepackage{color}
\usepackage[normalem]{ulem}

\journal{Physica A}
\newcommand\myFigureWidth{0.30}

\begin{document}

\begin{frontmatter}

\title{On absence of steady state in the Bouchaud-M{\'e}zard network model}

\author[mymainaddress]{Zhiyuan Liu}
\author[mymainaddress]{R. A. Serota\fnref{myfootnote}}
\fntext[myfootnote]{serota@ucmail.uc.edu}

\address[mymainaddress]{Department of Physics, University of Cincinnati, 
Cincinnati, Ohio 45221-0011}

\begin{abstract}
In the limit of infinite number of nodes (agents), the It{\^o}-reduced  
Bouchaud-M{\'e}zard network model of economic exchange has a time-independent 
mean and a steady-state inverse gamma distribution. We show that for a finite 
number of nodes the mean is actually distributed as a time-dependent lognormal 
and inverse gamma is quasi-stationary, with the time-dependent scale parameter. 
\end{abstract}


\end{frontmatter}

\section{Introduction\label{Introduction}}

Bouchaud-M{\'e}zard (BM) network model of economic exchange 
\cite{bouchaud2000wealth} sparked much interest in econophysics community 
because it organically predicted a power-law-tailed -- inverse gamma (IGa) -- 
distribution for the ``wealth". Recently, questions of relaxation in this model 
came to the forefront 
\cite{ma2014model,bouchaud2015growth,gabaix2016dynamics,berman2016far}. For constant mean 
wealth, relaxation times of the cumulants and of the entire distribution were 
studied in \cite{liu2017correlation}. Here we ascertain that for any finite 
number of nodes the mean should not be considered a constant and IGa a 
steady-state distribution. 

The fully connected BM network model \cite{bouchaud2000wealth} can be reduced to 
the following It{\^o} process for a node $i$ of the total of $M$ network nodes:
 \begin{equation}
\mathrm{d}w_i = \sqrt{2}\sigma w_i \textrm{d}B_i 
+ J(\overline{w}-w_i) \mathrm{d}t
\label{fpw2}
\end{equation}
where $J$ is the connection (exchange) strength (assumed equal between all 
nodes),  $B_i$ is a Wiener process, $\sigma$ is the strength of 
multiplicative stochasticity and
\begin{equation}
\overline{w}=\frac{1}{M} {\sum_{i}^{M} w_i}
\label{NetworkMean}
\end{equation}
is the mean over the network.

Our main result is that the quasi-stationary distribution function in a network 
is given by 
\begin{equation}
P(w) = \mathrm{IGa}(w;\, \alpha,\, \beta) = \frac{1}{w\Gamma(\alpha)} 
\left(\frac{\beta}{w}\right)^\alpha e^{-\frac{\beta}{w}}
\label{DistIGa}
\end{equation}
where $\alpha = 1 + J / \sigma^2$ is the shape parameter ($\alpha+1$ is the exponent of the 
power-law tail), $\beta(t) = \overline{w}(t) J / \sigma^2$ is the scale 
parameter and $\overline{w}(t)$ is 
the running mean distributed as lognormal (LN).
\begin{equation}
P(\overline{w},\,t) = \frac{1}{2 \sqrt{\pi  t} \sigma_M \overline{w}} \exp 
\left(-\frac{1}{2} \left(\frac{\log(\overline{w}) + \sigma_M^2 t}{\sqrt{2 
t}\sigma_M}\right)^2\right)
\label{DistLN}
\end{equation}
\begin{equation}
\sigma_M = \frac{\sigma}{\sqrt{M}}\sqrt{\frac{\alpha-1}{\alpha-2}}
\label{sigmaMIGa}
\end{equation}
where the distribution is understood as over a system of BM networks. The mean 
and the variance of the mean are then respectively
\begin{equation}
\langle \overline{w} \rangle =1
\label{OmegaMean}
\end{equation}
\begin{equation}
\langle \overline{w}^2 \rangle - \langle \overline{w} \rangle^2 
=\exp\left(2\sigma_M^2t\right) - 1
\label{OmegaVar}
\end{equation}
Unity in (\ref{OmegaMean}) is due to the fact that in the corresponding 
infinite-node case $\overline{w}$ in (\ref{fpw2}) and (\ref{DistIGa}) is 
constant and can be set to unity by simple rescaling. In our calculations and simulations we use all $w_i=1$ as the initial condition.

This paper is organized as follows.
Secs. \ref{MeanDistVar} and \ref{NumSim} below contain respectively the 
analytical derivation and its numerical verification. We conclude with the 
discussion of possible extensions of our work.

\section{Analytical derivation\label{MeanDistVar}}

Here we first give a simple derivation of the LN distribution of the mean and 
then provide a detailed analytical derivation of the variance of the mean that 
independently verifies the LN result.
  
\subsection{Distribution of the mean\label{ConnNet}}

For the BM network (\ref{fpw2}), from Eqs. (\ref{fpw2}) and (\ref{NetworkMean})
\begin{equation}
\mathrm{d}\overline{w} = \frac{\sqrt{2}\sigma}{M}\sum_{i=1}^{M} w_i 
\textrm{d}B_i 
\label{fpwmLN}
\end{equation}
that is the exchange term is eliminated. Notice, that this means that 
(\ref{fpwmLN}) is formally the same as for the completely disconnected network 
(see {\ref{AppendixLN}}). Surmising that the distribution of the mean for the 
BM network is LN, we seek to replace the r.h.s. of (\ref{fpwmLN}) with
\begin{equation}
\mathrm{d}\overline{w} = \sqrt{2}\sigma_M\overline{w}\textrm{d}B
\label{fpwmLN0}
\end{equation}
Then squaring and equating the r.h.s. of Eqs. (\ref{fpwmLN}) and 
(\ref{fpwmLN0}), with the off-diagonal terms being zero, we obtain
\begin{equation}
\sigma_M^2= \frac{\sigma^2}{M^2\overline{w}^2}\sum_{i=1}^{M} w_i^2
\label{A}
\end{equation}

To determine $\sigma_M$ we made a crucial assumption -- which will be confirmed 
numerically in Sec. \ref{NumSim} -- that the quasi-stationary distribution in a 
network is given by (\ref{DistIGa}). Then replacing 
$\overline{w}=\left(\sum_{i}^{M} w_i \right) / M$ and $\left(\sum_{i=1}^{M} 
w_i^2\right)/M$ with the expectation values of $w$ and $w^2$ computed with the 
distribution (\ref{DistIGa}), we obtain
\begin{equation}
\sigma_M = \frac{\sigma}{\sqrt{M}}\sqrt{\frac{\alpha-1}{\alpha-2}}
\label{sigmaMIGa2}
\end{equation}
Since the resultant $\sigma_M$ is time-independent, by It{\^o} 
calculus (\ref{fpwmLN0}) becomes
\begin{equation}
\mathrm{d}\ln \overline{w} = \sqrt{2}\sigma_M \mathrm{d}B - \sigma_M^2 
\mathrm{d}t
\end{equation}
which yields the LN distribution (\ref{DistLN}).

\subsection{Variance}\label{Variance}

Denoting $\kappa_2$ the variance of the mean $\overline{w}$ and setting $\langle 
\overline{w}\rangle = 1$, we have
\begin{equation}
\label{kappa2}
\kappa_2 = \langle\overline{w}^2\rangle - \langle\overline{w}\rangle^2 = 
\langle\overline{w}^2\rangle - 1
\end{equation}
Then, using (\ref{fpwmLN}) and that $\left\langle \overline{w} \cdot 
\overline{w_i \mathrm{d}B_i} \right\rangle = 0$, we obtain
\begin{equation}
\begin{split}
\mathrm{d}\overline{w}^2 & = 2\overline{w}\mathrm{d}\overline{w} + 
(\mathrm{d}\overline{w})^2 \\
& = 2\overline{w} \cdot \sqrt{2}\sigma \frac{1}{M}\sum_{i=1}^{M} w_i 
\textrm{d}B_i + \left( \sqrt{2}\sigma \frac{1}{M}\sum_{i=1}^{M} w_i 
\textrm{d}B_i \right)^2 \\
& = 2\sqrt{2}\sigma\overline{w} \cdot \overline{w_i \mathrm{d}B_i} + 2\sigma^2 
\overline{w_i \mathrm{d}B_i}^2
\end{split}
\end{equation}
and
\begin{equation}
\mathrm{d}\kappa_2 = 2\sigma^2 \left\langle \overline{w_i \mathrm{d}B_i}^2 
\right\rangle=2\sigma^2 \left\langle \left(\frac{1}{M}\sum_{i=1}^{M} w_i 
\mathrm{d}B_i \right)^2 \right\rangle
\end{equation}
Since,
\begin{equation}
\begin{split}
\left( \frac{1}{M}\sum_{i=1}^{M} w_i \mathrm{d}B_i \right)^2 & = \frac{1}{M^2} 
\sum_{i=1}^{M} w_i^2 (\mathrm{d}B_i)^2 + \frac{1}{M^2} \sum_{i=1}^{M} \sum_{j=1, 
j\neq i}^{M} w_i w_j \mathrm{d}B_i \mathrm{d}B_j
\end{split}
\end{equation}
then averaging, replacing $(\mathrm{d}B_i)^2 = \mathrm{d}t$, and eliminating the 
off-diagonal terms due to $\mathrm{d}B_i$ and $\mathrm{d}B_j$ being independent, 
we find
\begin{equation}
\label{dkappa2}
\mathrm{d}\kappa_2 = \frac{2\sigma^2}{M^2} \left\langle \sum_{i=1}^{M} w_i^2 
\right\rangle \mathrm{d}t = \frac{2\sigma^2}{M} \left\langle \frac{1}{M} 
\sum_{i=1}^{M} w_i^2 \right\rangle \mathrm{d}t
\end{equation}
Notice now that
\begin{equation}
\label{moment}
\left\langle \frac{1}{M}\sum_{i=1}^{M} w_i^2 \right\rangle = \left\langle 
\frac{1}{M}\sum_{i=1}^{M} \left( w_i - \overline{w} \right)^2 \right\rangle + 
\langle\overline{w}^2\rangle
\end{equation}
Per our surmise (\ref{DistIGa}), the variance in each path is 
$\overline{w}^2/(\alpha - 2)$, so that (\ref{moment}) yields per 
(\ref{kappa2})
\begin{equation}
\left\langle \frac{1}{M}\sum_{i=1}^{M} w_i^2 \right\rangle =  \frac{\langle 
\overline{w}^2 \rangle}{\alpha - 2} + \langle\overline{w}^2\rangle = 
\frac{\alpha - 1}{\alpha - 2} \cdot \left(\kappa_2 + 1\right) 
\end{equation}
which reduces (\ref{dkappa2}), with the use of (\ref{sigmaMIGa}), to
\begin{equation}
\mathrm{d} \kappa_2 = 2 \sigma_M^2 \left(\kappa_2 + 1\right)  \mathrm{d}t
\end{equation}
Solving the differential equation with the initial condition of $\kappa_2 = 0$ 
when $t = 0$, we finally obtain (\ref{OmegaVar})
\begin{equation}
\kappa_2 = \exp\left(2\sigma_M^2 t\right) - 1
\label{kappa_2IGaFinal}
\end{equation}

\section{Numerical simulations}\label{NumSim}

We simulate (\ref{fpw2}) using $J=0.1$, $\sigma^2=0.05$ and $M=2^{13}$. The time 
step is $\mathrm{d}t=2^{-6}$, that is we use $2^6$ steps between two time 
``ticks", $\Delta t=1$. Since in the infinite-node BM system the fixed mean value 
can be always set to unity, we use $w_i=1$ as the initial value for all nodes in 
all networks. For averaging over and for distribution over the networks, we use 
$2^{11}$ networks. 

\begin{figure}[!htbp]
\centering
\begin{tabular}{ccc}
\includegraphics[width = \myFigureWidth \textwidth]{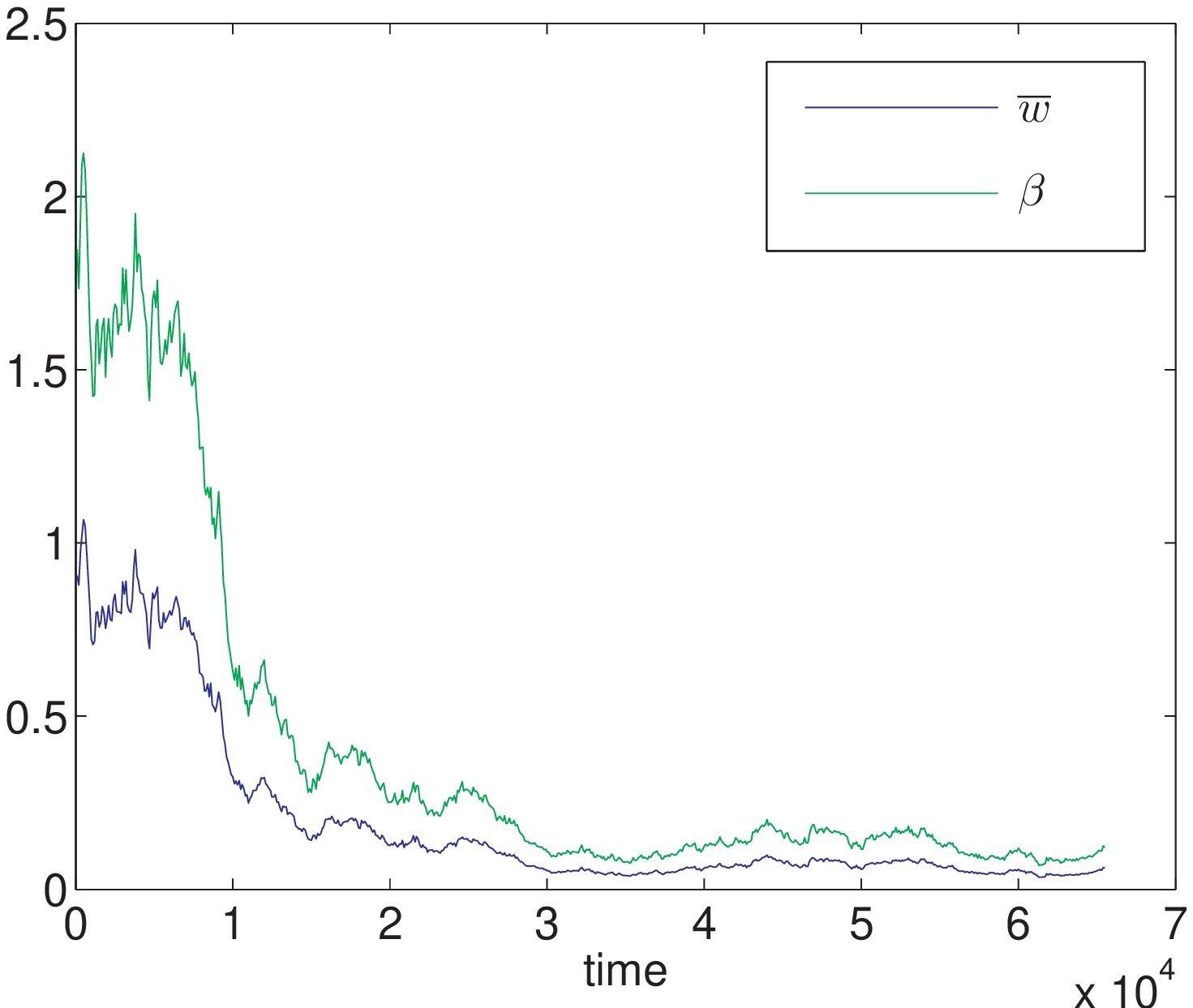} &
\includegraphics[width = \myFigureWidth \textwidth]{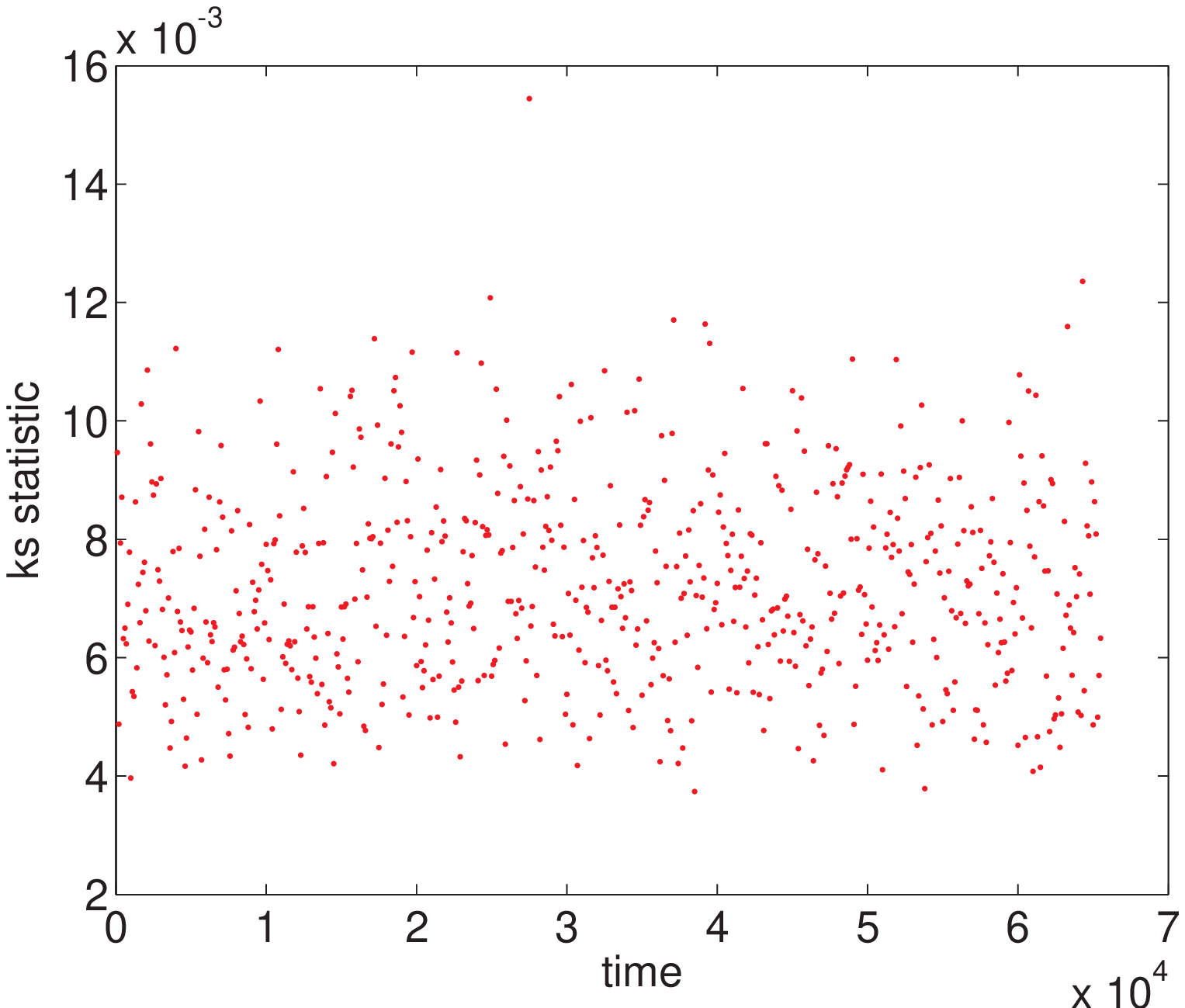} &
\includegraphics[width = \myFigureWidth \textwidth]{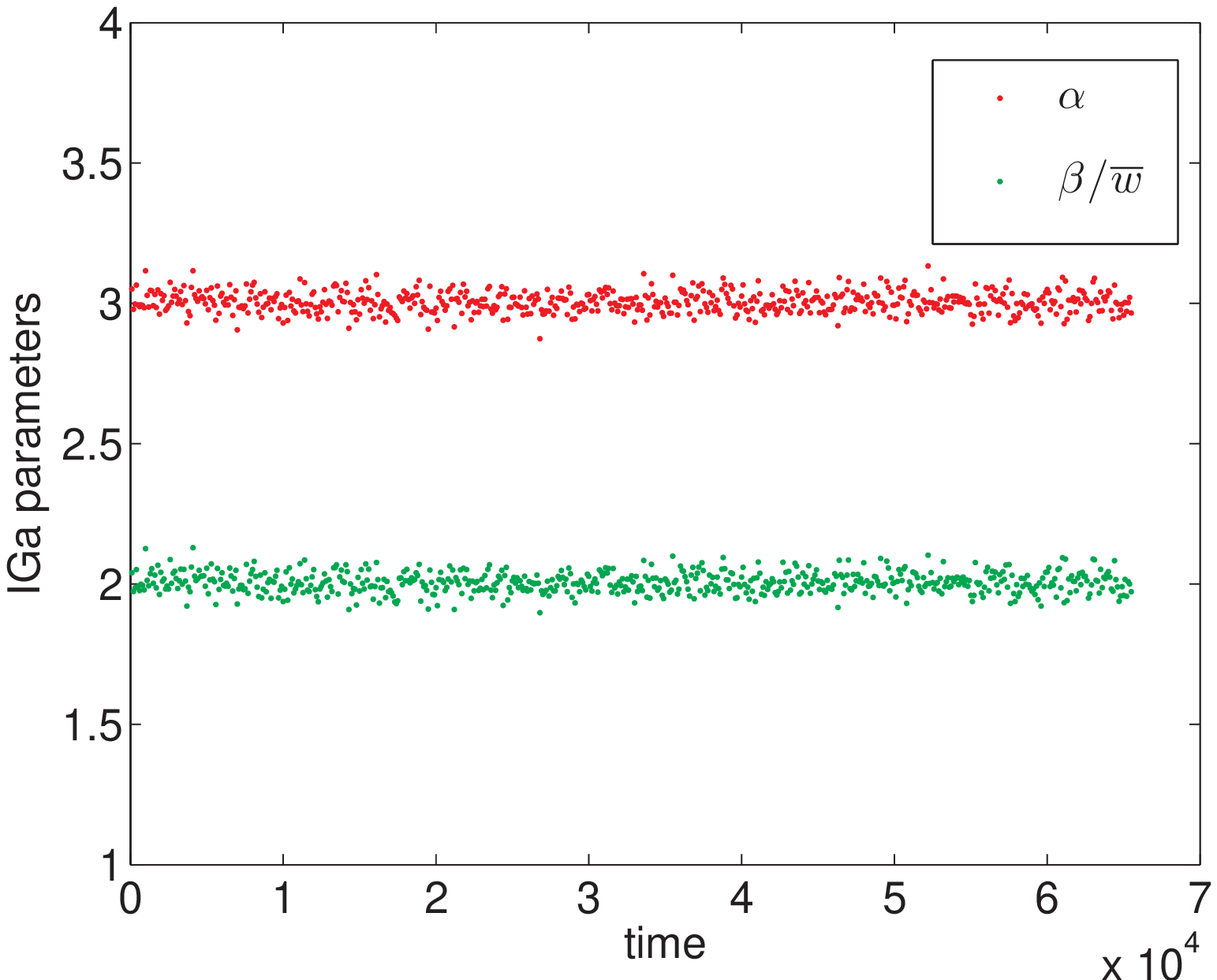} \\
\includegraphics[width = \myFigureWidth \textwidth]{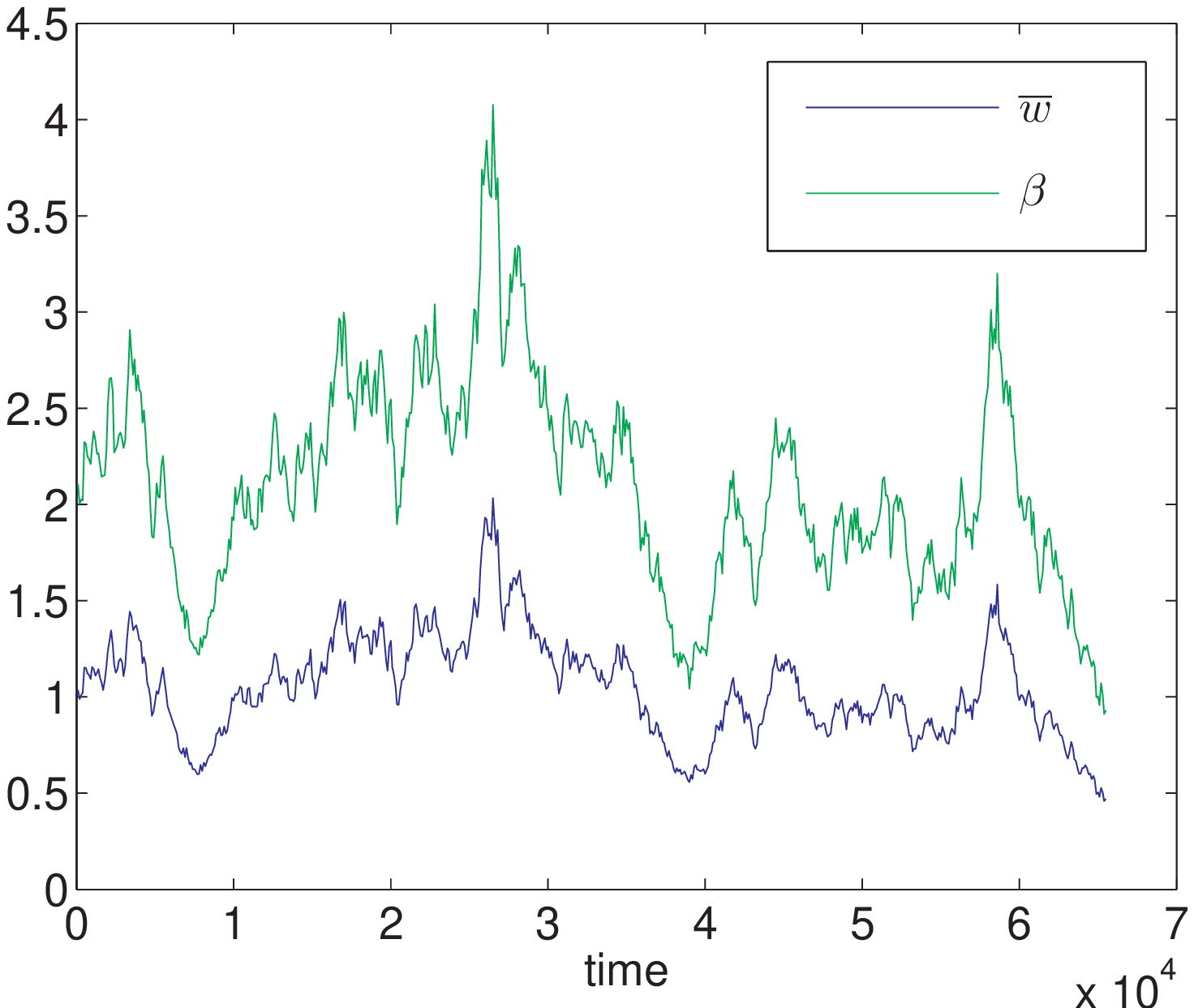} &
\includegraphics[width = \myFigureWidth \textwidth]{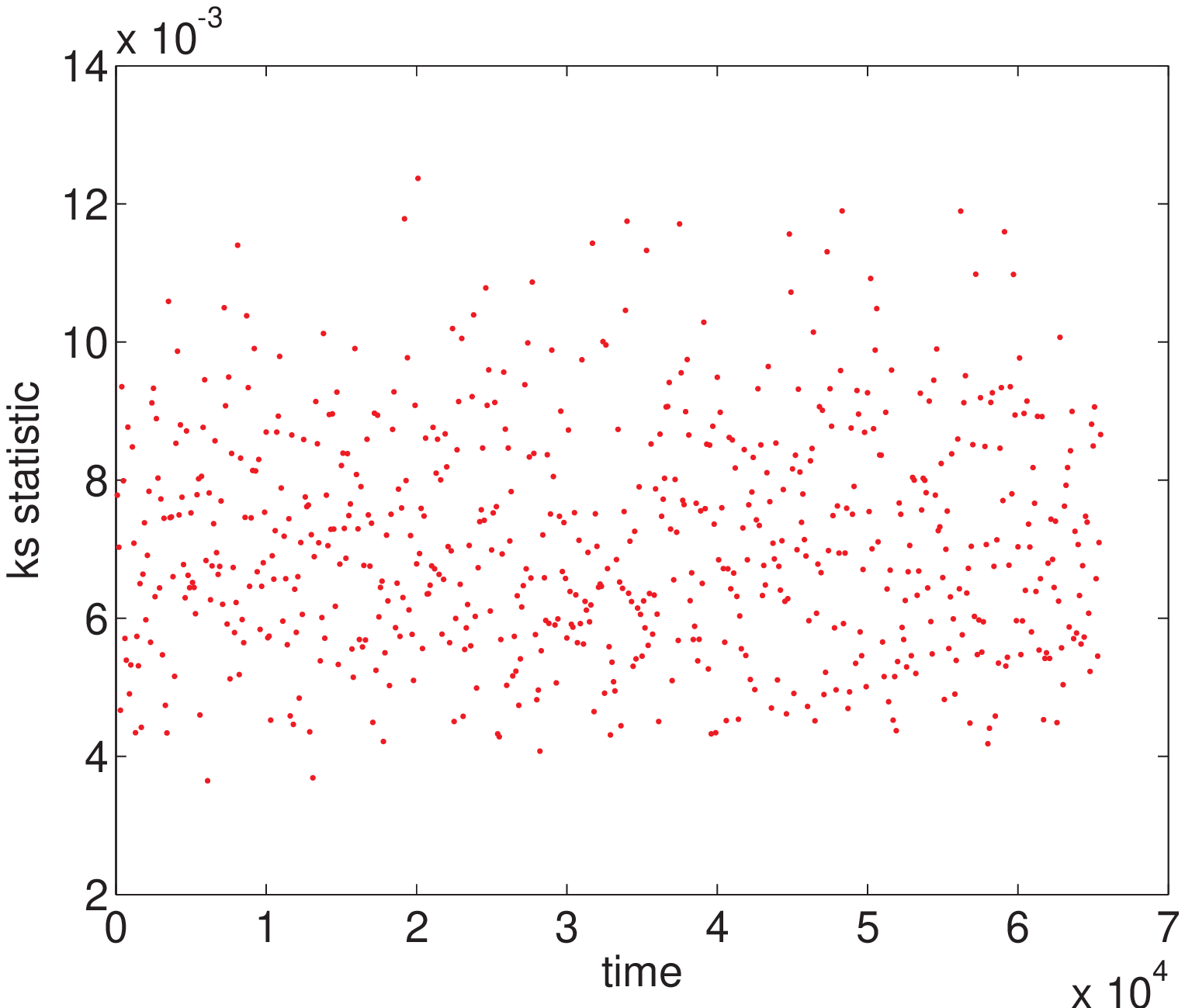} &
\includegraphics[width = \myFigureWidth \textwidth]{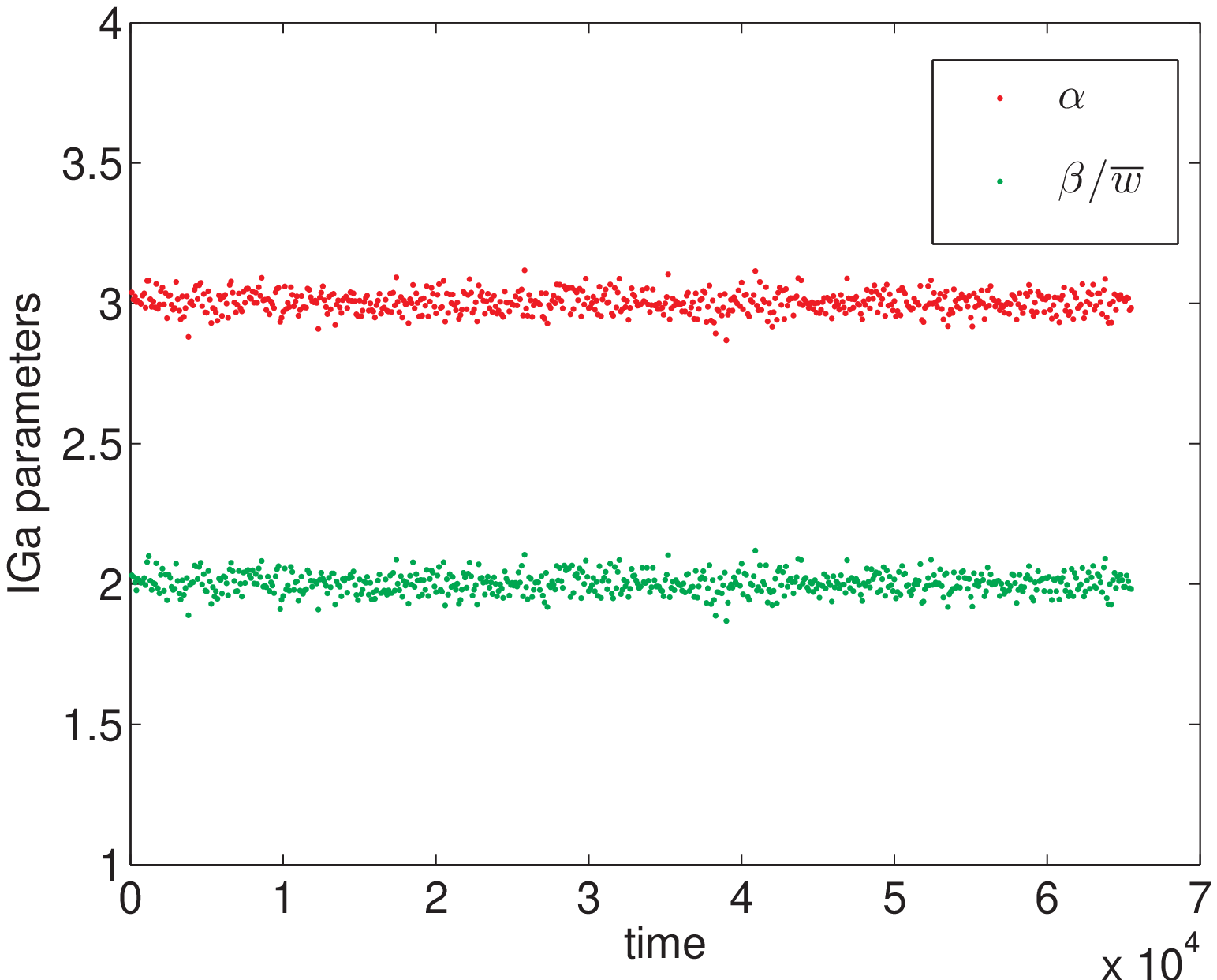} \\
\includegraphics[width = \myFigureWidth \textwidth]{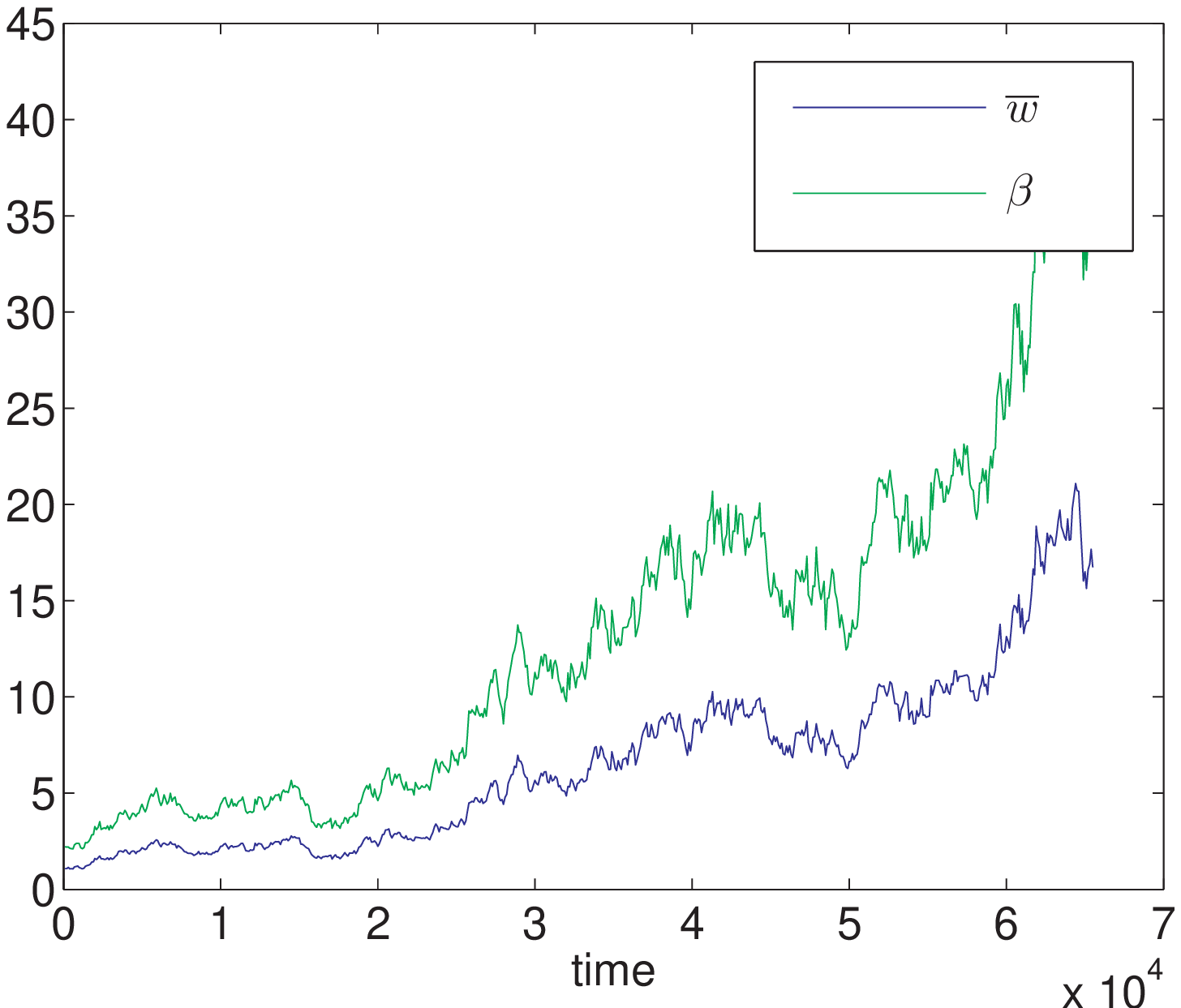} &
\includegraphics[width = \myFigureWidth \textwidth]{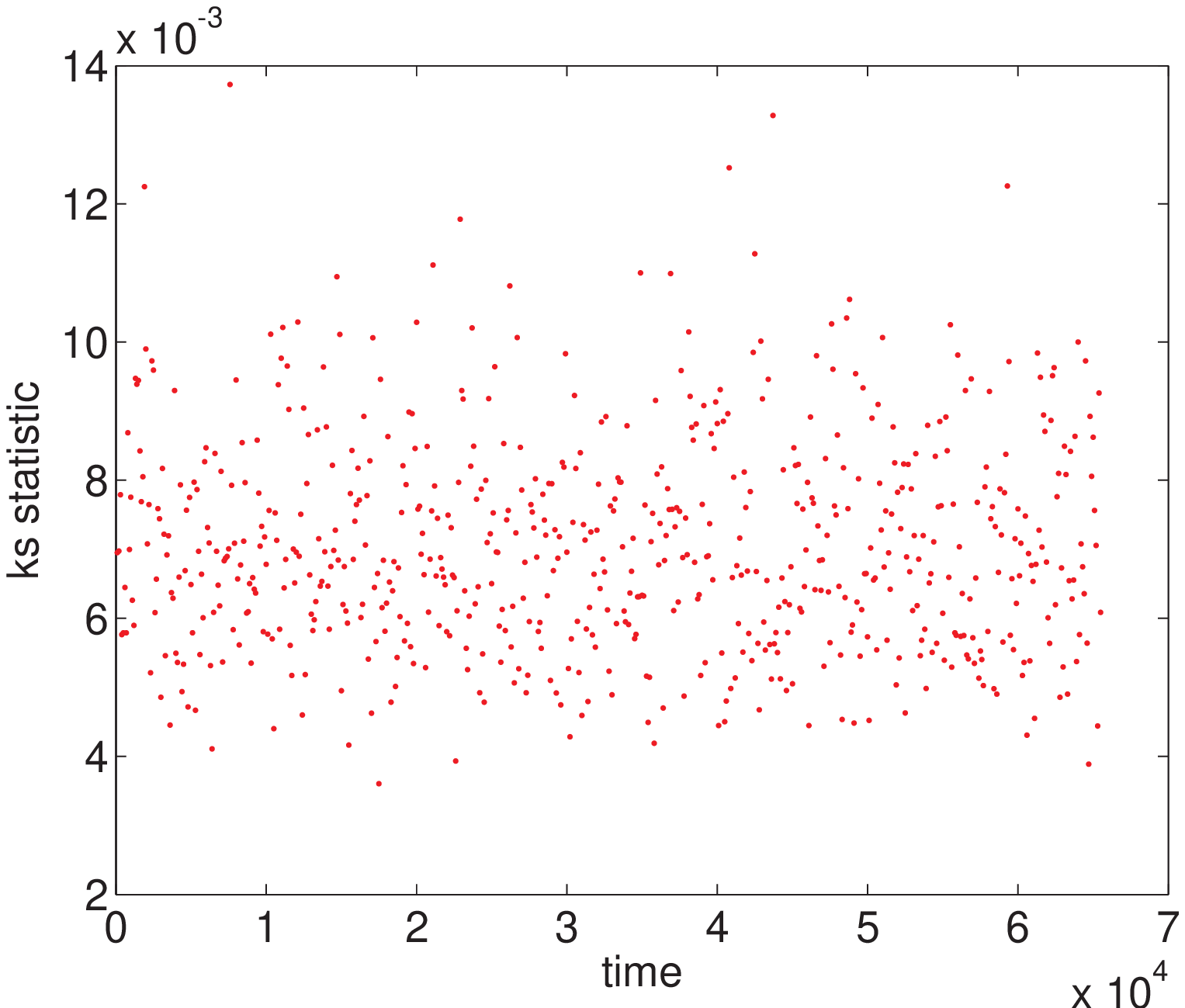} &
\includegraphics[width = \myFigureWidth \textwidth]{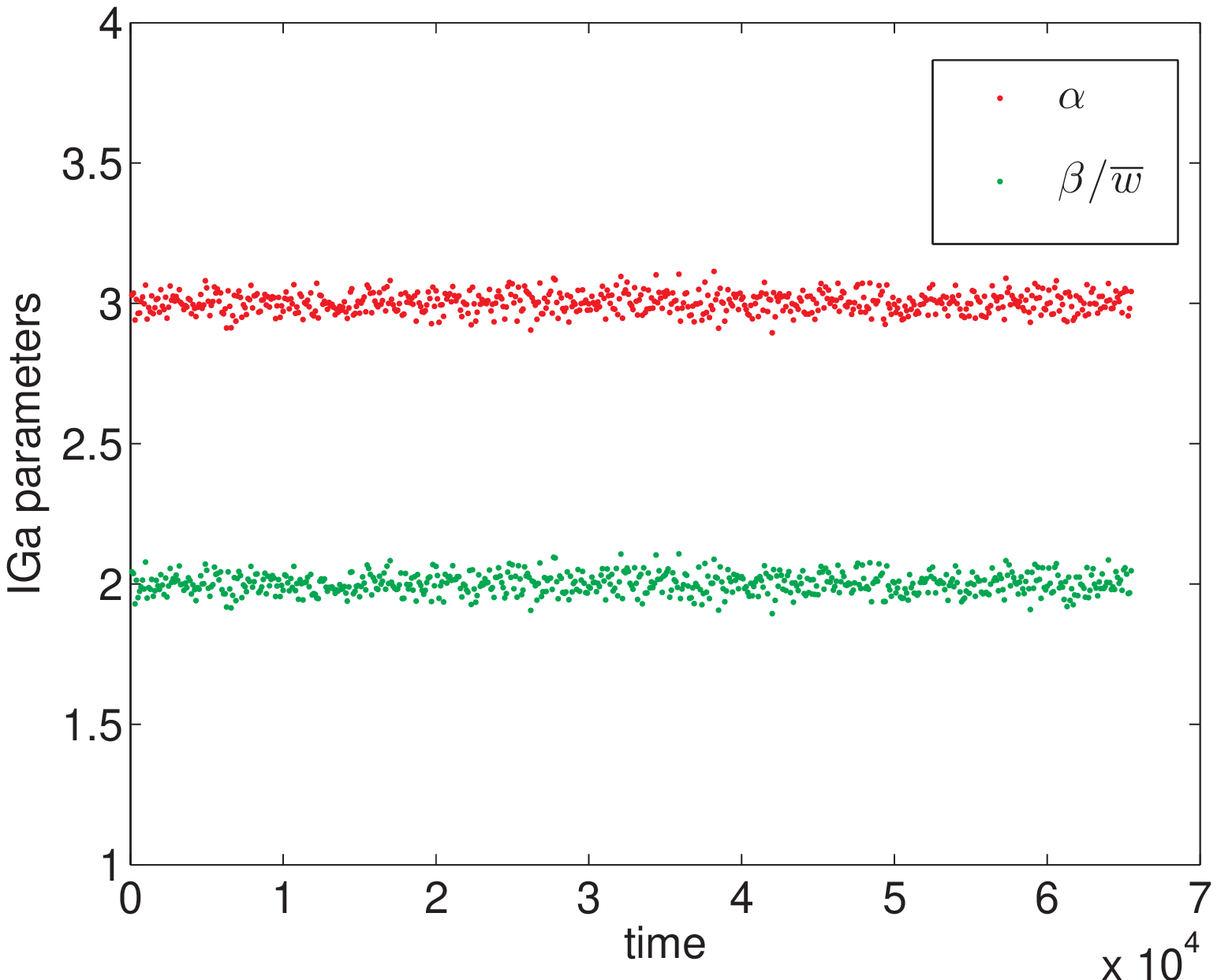} \\
\end{tabular}
\caption{Each row corresponds to a different network. Left column: $\beta$ -- 
top (green) line and $\overline{w}$ -- bottom (blue) line; Middle column -- KS 
statistic for fitting with IGa distribution; Right column: parameters $\alpha$ 
-- top (red) line and $\beta/\overline{w}$ -- bottom (green) line.}
\label{fullyConnectedNumericalResults}
\end{figure}

The left column in Fig. \ref{fullyConnectedNumericalResults} shows segments of 
typical behavior of the running mean: increasing, oscillating around unity and 
decreasing. The middle column shows Kolmogorov-Smirnov (KS) statistic for 
fitting with IGa and the right column the parameters of the fitted IGa. The 
results are clearly in excellent agreement with our surmise (\ref{DistIGa}).

\begin{figure}[!htbp]
\centering
\includegraphics[width = \myFigureWidth \textwidth]{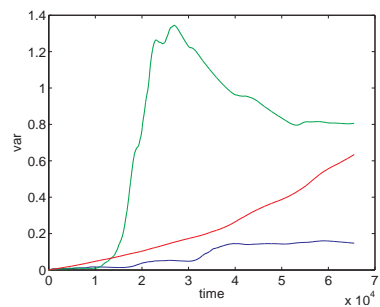}
\caption{Variance of the mean in a single network for two different networks -- 
top (green) and bottom (blue) lines; Average over the networks -- middle (red) 
line.}
\label{testtestAverageSDWithTime00001}
\end{figure}

The top (green) and bottom (blue) in Fig. (\ref{testtestAverageSDWithTime00001}) 
are examples of the variance of the running mean $\overline{w}(t)$ in a single 
network.The middle (red) line is the average over all networks.

\begin{figure}[!htbp]
\centering
\includegraphics[width = \myFigureWidth \textwidth]{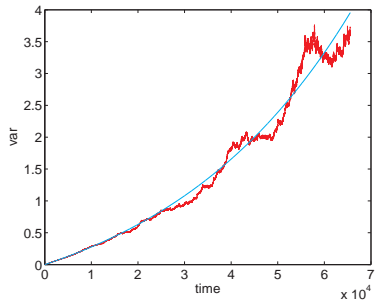}
\caption{Variance of the instantaneous $\overline{w}(t)$ over the networks: 
theoretical result (\ref{kappa_2IGaFinal}) -- smooth (cyan) line; numerical 
simulation -- jagged (red) line.}
\label{varianceOverPaths}
\end{figure}

Fig. (\ref{varianceOverPaths}) is the test of Eq. (\ref{kappa_2IGaFinal}). At a 
given time $t$, we evaluate the variance of $\overline{w}(t)$ over the networks. 
The smooth (cyan) line is the theoretical result and the jagged (red) line is 
the numerical simulation.

\begin{figure}[!htbp]
\centering
\includegraphics[width = \myFigureWidth \textwidth]{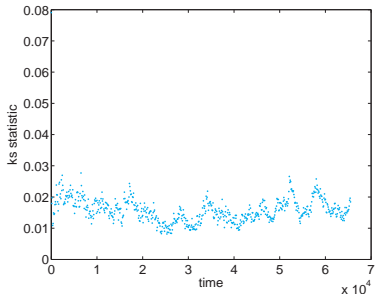}
\caption{KS statistic as a function of time for fitting instantaneous 
$\overline{w}(t)$ over networks with LN.}
\label{compareKSTest}
\end{figure}

Finally, Fig. \ref{compareKSTest} shows the KS statistic for fitting the 
distribution of $\overline{w}(t)$ over the networks with LN as a function of 
time.

\section{Conclusion}
We explored relaxation in the Bouchaud-M{\'e}zard network model and showed that 
for any finite number of nodes there is no steady state per se. At long times, 
the distribution of node values is described by a quasi-stationary IGa
distribution with the fixed shape parameter, that defines the exponent of the 
power-law tail, and variable scale parameter proportional to the running mean. 
The distribution of the running mean over a system of Bouchaud-M{\'e}zard 
networks is $\textrm{LN}(\overline{w};\, -Q/2,\sqrt{Q})$, where $Q$ is linear in time and 
inversely proportional to the number of nodes. For the infinite number of nodes, 
the problem reduces to the standard IGa process, whose relaxation to 
the IGa distribution with the constant mean (which can always be set 
to unity) had been previously explored \cite{liu2017correlation}. 

An obvious extension of this work is to the partially connected networks -- 
random and regular -- which are described by the generalized inverse gamma 
distribution \cite{ma2013distribution}. We have numerical results that parallel our findings for the fully 
connected networks, namely, that we have a quasi-stationary generalized inverse 
gamma distribution and a time-dependent heavy-tailed distribution of the mean 
(possibly also lognormal). However, we do not yet have a clear path to 
analytical derivation. Extension to relaxation processes in other stochastic 
processes believed to have steady-state distributions is another obvious 
direction of this research.

\appendix
\section{Disconnected network}\label{AppendixLN}
A disconnected network allows for a an independent check of the formalism 
developed in Sec. \ref{Variance}. In a disconnected network, $J = 0$,
\begin{equation}
\mathrm{d}w_i = \sqrt{2}\sigma w_i\textrm{d}B_i 
\label{fpwLN}
\end{equation}
we have a LN distribution for each node
\begin{equation}
P(w,\,t) = \frac{1}{2 \sqrt{\pi  t} \sigma w} \exp \left(-\frac{1}{2} 
\left(\frac{\log(w) + \sigma^2 t}{\sqrt{2 t}\sigma}\right)^2\right)
\label{DistLN1}
\end{equation}
with average and variance given by (\ref{OmegaMean}) and (\ref{OmegaVar}), where 
$\sigma_M$ is replaced with $\sigma$.

Since all nodes are independent, the distribution of the network mean (\ref{NetworkMean}) is determined by the average
of independent and identically distributed (i.i.d.) LN 
random variables. Therefore, per the central limit theorem (CLT), one expects 
the variance of the mean to be $\approx \exp{\left(2 \sigma^2t\right)}/M$ for 
large $M$. All the steps through (\ref{moment}) in Sec. \ref{Variance} are the 
same here while the rest are now as follows:
\begin{equation}
\left\langle \frac{1}{M}\sum_{i=1}^{M} w_i^2 \right\rangle = \left(e^{2 
\sigma^2t}-1\right) + \left(\kappa_2 + 1\right) =e^{2 \sigma^2t}+\kappa_2
\end{equation}
\begin{equation}
\mathrm{d} \kappa_2 = \frac{2\sigma^2}{M} \left(\kappa_2 + e^{2 
\sigma^2t}\right)  \mathrm{d}t
\end{equation}
which indeed yields 
\begin{equation}
\kappa_2 =  \frac{1}{M-1} \left(e^{2 \sigma^2t}-e^{\frac{2 \sigma^2t}{M}}\right) 
\approx \frac{e^{2 \sigma^2t}}{M}
\end{equation}
for large $M$.

The sum of LN distributions can be well approximated as LN distribution 
\cite{mehta2007approximating}, so it is tempting to think that the average of $M$ LN 
distributions may be approximated by a $\sqrt{M}$-narrower LN distribution, 
which tends to the normal (N) distribution while maintaining the character of 
the LN heavy tail \cite{lam2011corrections}. (Incidentally, the sum of 
i.i.d. IGa random variables is well approximated by a IGa distribution so it 
is tempting again to approximate the average of $M$ IGa distributions as a
a $\sqrt{M}$-narrower IGa distribution, tending to N while maintaining the 
character of the IGa fat tail \cite{dashti2017untitled}. Of course, in the BM 
model the variables in the sum for the network mean are not independent.)
A lognormal $\textrm{LN}(w;\, -Q/2,\, \sqrt{Q})$, where
$Q = \log \left(\kappa_2 + 1\right)$, has the required variance and unity mean.
In the large $M$ limit, its explicit form is as follows:
\begin{equation}
P(w,\,t) = \frac{1} {\sqrt{2 \pi  \log \left(\frac{M+e^{2 \sigma ^2 t}}{M}\right)}
w} \exp \left(-\frac{1}{2}\frac{\left( \log (w)+\frac{1}{2}\log \left(\frac{M 
+ e^{2 \sigma ^2 t}}{M}\right)\right)^2}{ \log \left(\frac{M+e^{2 
\sigma ^2 t}}{M}\right)}\right)
\end{equation}

\bibliography{mybib}

\end{document}